\documentclass[twocolumn,showpacs,showkeys,amsmath,amssymb,superscriptaddress,floatfix,nofootinbib]{revtex4}

\usepackage{graphicx}
\usepackage{bm} 
\usepackage{subfigure}
\def\vec#1{\mathchoice{\mbox{\boldmath$\displaystyle#1$}}
{\mbox{\boldmath$\textstyle#1$}}
{\mbox{\boldmath$\scriptstyle#1$}}
{\mbox{\boldmath$\scriptscriptstyle#1$}}}
\makeatletter
\newcommand\erfc{\mathop{\operator@font erfc}\nolimits}
\def\slashchar#1{\setbox0=\hbox{$#1$}
   \dimen0=\wd0 \setbox1=\hbox{/} \dimen1=\wd1
   \ifdim\dimen0>\dimen1 \rlap{\hbox to \dimen0{\hfil/\hfil}} #1
   \else  \rlap{\hbox to \dimen1{\hfil$#1$\hfil}} / \fi}

\begin{document}
 
\title{
Early anisotropic hydrodynamics and the RHIC early-thermalization and HBT puzzles}

\author{Radoslaw Ryblewski} 
\affiliation{The H. Niewodnicza\'nski Institute of Nuclear Physics, Polish Academy of Sciences, PL-31342 Krak\'ow, Poland}

\author{Wojciech Florkowski} 
\affiliation{Institute of Physics, Jan Kochanowski University, PL-25406~Kielce, Poland} 
\affiliation{The H. Niewodnicza\'nski Institute of Nuclear Physics, Polish Academy of Sciences, PL-31342 Krak\'ow, Poland}

\date{April 9, 2010}

\begin{abstract}
We address the problem if the early thermalization and HBT puzzles in relativistic heavy-ion collisions may be solved by the assumption that the early dynamics of the produced matter is locally anisotropic. The hybrid model describing the purely transverse hydrodynamic evolution followed by the perfect-fluid hydrodynamic stage is constructed. The transition from the transverse to perfect-fluid hydrodynamics is described by the Landau matching conditions applied at a fixed proper time $\tau_{\rm tr}$. The global fit to the RHIC data reproduces the soft hadronic observables (the pion, kaon, and the proton spectra, the pion and kaon elliptic flow, and the pion HBT radii) with the accuracy of about 20\%. These results indicate that the assumption of the very fast thermalization may be relaxed. In addition, the presented model suggests that a large part of the inconsistencies between the theoretical and experimental HBT results may be removed. 
\end{abstract}



\maketitle 


\section{Introduction}
\label{sect:intro}

The experimental results obtained at the Relativistic Heavy-Ion Collider (RHIC) are nowadays interpreted as the evidence that the matter produced in relativistic heavy-ion collisions equilibrates very fast (presumably within a small fraction of 1~fm/c) and its behavior is very well described by the perfect-fluid hydrodynamics  \cite{Kolb:2003dz,Huovinen:2003fa,Shuryak:2004cy,Teaney:2001av,Hama:2005dz,Hirano:2007xd,Nonaka:2006yn,Bozek:2009ty}. The fast equilibration and perfect-fluidity are naturally explained by the assumption that the produced matter is a strongly coupled quark-gluon plasma (sQGP) \cite{Shuryak:2004kh}. Another possible explanation assumes that the plasma is weakly interacting, however the plasma instabilities lead to the fast isotropization of matter, which in turn helps to achieve equilibration \cite{Mrowczynski:2005ki}. 

The concept of a very fast equilibration is difficult to reconcile with the models of the early stages of the relativistic heavy-ion collisions, which typically use the ideas of color strings or color-flux tubes. The system produced by strings is highly anisotropic; the pressure in the direction transverse to the collision axis is usually much larger than the longitudinal pressure. 

Interestingly, the decays of fluctuating strings produce matter that looks as if it was  thermalized only in the transverse direction, i.e., the transverse-momentum spectra of partons are exponential \cite{Bialas:1999zg}. This feature inspired the construction of the hydrodynamic model where only transverse degrees of freedom are thermalized, while the longitudinal motion is described by the free streaming \cite{Bialas:2007gn}. We shall call this framework the {\it transverse hydrodynamics}. The first calculations using this idea showed that it was consistent with the data describing the transverse-momentum spectra and the elliptic flow coefficient $v_2$ of the pions measured at RHIC \cite{Bialas:2007gn,Chojnacki:2007fi}. Further studies, where the transverse hydrodynamics was combined with the statistical hadronization model, showed that one may reproduce in this framework also the HBT pion radii in a quite successful way \cite{Ryblewski:2009hm,Florkowski:2009wb}. 

The concept of transverse hydrodynamics allows us to avoid the problem of early thermalization. Nevertheless, the RHIC data suggests that the matter is eventually fully thermalized (by this we mean that the {\it local thermodynamic equilibrium} with the locally isotropic momentum distribution is achieved). Thus, we expect that during the evolution of the system the early transverse-hydrodynamics stage should change into the stage described by the perfect-fluid hydrodynamics (with possible small corrections from viscosity and other dissipative effects). In this paper we present the hybrid model that incorporates such a change. In this way we generalize the methods and concepts introduced earlier in Refs. \cite{Bialas:2007gn,Chojnacki:2007fi,Ryblewski:2009hm,Florkowski:2009wb}

We assume that the transverse-hydrodynamics stage is valid in the proper time interval $\tau_{\rm i} \leq \tau \leq \tau_{\rm tr} $. The proper time $\tau_{\rm i}$ defines the initialization time for transverse hydrodynamics, and $\tau_{\rm tr}$ is the transition time from the transverse hydrodynamics to the perfect-fluid hydrodynamics~\footnote{We shall refer to this transition below shortly as to the \mbox{2D $\to$ 3D transition}.}. This transition is described with the help of the Landau matching conditions. Both $\tau_{\rm i}$ and $\tau_{\rm tr} $ are the parameters of the model. For $\tau > \tau_{\rm tr} $ the system is treated as the perfect fluid. The freeze-out takes place on the hypersurface of constant temperature, and the Monte-Carlo method is used to calculate the soft hadronic observables: the transverse-momentum spectra of pions, kaons, and protons, the elliptic flow of pions+kaons and protons, and the HBT radii of pions.

As stated above, the idea of transverse hydrodynamics helps to solve the problem of very fast thermalization. There exists, however, another important problem in the interpretation of the RHIC data, namely, the HBT puzzle. This name refers to general problems connected with the correct  reproduction of the HBT radii in the hydrodynamic models. One possible solution of this problem suggests the use of the {\it modified initial conditions} for the perfect-fluid evolution \cite{Broniowski:2008vp}  --- instead of the initial energy density profiles in the transverse plane obtained from the Glauber model one should use the Gaussian profiles (with the same widths as in the Glauber calculation). 

In this paper we use the standard Glauber initial conditions and check how our {\it modified dynamics} (i.e., the dynamics including  the transverse-hydrodynamics stage) affects the HBT radii. We find that our model offers a good description of the HBT radii which are reproduced at the level of about 10\%. The pion spectra and $v_2$ are described with the accuracy of about 10\% and 20\%, respectively. This means that our approach provides a very good description of the pion production at RHIC. 

The global fit to the RHIC data (including kaons and protons) reproduces the soft hadronic observables with the accuracy of about 20\%. The worst model result are obtained for the proton elliptic flow, which is 50\% larger than the measured value. In our opinion this is caused by the neglecting of the hadronic rescattering phase (or the effects of the shear and bulk viscosities as shown in \cite{Bozek:2009dw}). In any case, the overall good description of the data obtained in our model is encouraging, especially from the point of view that the problem of early thermalization is circumvented and the inconsistencies between the theoretical and experimental HBT results are strongly reduced. 

\bigskip

Throughout the paper we use the three-dimensional (3D) and two-dimensional (2D) densities of various physical quantities. Thus, to avoid possible confusion we have introduced the subscripts ``3'' or ``2''. For example, the 3D energy density (energy per unit volume) is denoted by $\varepsilon_{3}$, while the 2D energy density (energy per unit area in the transverse plane) is denoted by $\varepsilon_{2}$.

The rapidity and spacetime rapidity are defined by the expressions
\begin{eqnarray}
y = \frac{1}{2} \ln \frac{E+p_\parallel}{E-p_\parallel}, \quad
\eta = \frac{1}{2} \ln \frac{t+z}{t-z}, \label{yandeta} 
\end{eqnarray}
which follow from the standard parameterization of the four-momentum and spacetime coordinate of a particle,
\begin{eqnarray}
p^\mu &=& \left(E, {\vec p}_\perp, p_\parallel \right) =
\left(m_\perp \cosh y, {\vec p}_\perp, m_\perp \sinh y \right), \nonumber \\
x^\mu &=& \left( t, {\vec x}_\perp, z \right) =
\left(\tau \cosh \eta, {\vec x}_\perp, \tau \sinh \eta \right). \label{pandx}
\end{eqnarray} 
In Eq. (\ref{pandx}) the quantity $m_\perp$ is the transverse mass
\begin{equation}
m_\perp = \sqrt{m^2 + p_x^2 + p_y^2},
\label{energy}
\end{equation}
and $\tau$ is the proper time
\begin{equation}
\tau = \sqrt{t^2 - z^2}.
\label{tau}
\end{equation} 
Everywhere we use the natural units where $c=1$ and $\hbar=1$.

\section{Initial conditions}
\label{sect:init-cond}

\subsection{Initial transverse profiles}
\label{sect:init-prof}

For the boost-invariant systems, which are studied below, one usually assumes that either the initial entropy density, $\sigma_{3\rm i} ({\bf x}_\perp) = \sigma_3(\tau_{\rm i},{\bf x}_\perp)$, or the initial energy density, $\varepsilon_{3\rm i} ({\bf x}_\perp)=\varepsilon_3(\tau_{\rm i},{\bf x}_\perp)$, are directly related to the density of {\it sources of particle production}, $\rho_{\rm sr} ({\bf x}_\perp)$. These sources are identified with {\it wounded nucleons} or {\it binary collisions} \cite{Kolb:2001qz}. The symmetry with respect to the Lorentz boosts along the collision axis means that it is sufficient to consider all these quantities in the plane $z=0$. 

More generally, a mixed model is used \cite{Kharzeev:2000ph}, with a linear combination of the wounded-nucleon density ${\overline w}\left( {\bf x}_\perp \right)$ and the density of binary collisions ${\overline n} \left( {\bf x}_\perp \right)$. This leads to the two popular choices:
\begin{equation}
\sigma_{3\rm i} ({\bf x}_\perp) \;\propto\; \rho_{\rm sr} ({\bf x}_\perp) = 
\frac{1-\kappa}{2}\, {\overline w}\left( {\bf x}_\perp \right) 
+ \kappa\, {\overline n} \left( {\bf x}_\perp \right)
\label{eqn:initial_entropy}
\end{equation}
or
\begin{equation}
\varepsilon_{3\rm i} ({\bf x}_\perp) \;\propto\; \rho_{\rm sr} ({\bf x}_\perp) = 
\frac{1-\kappa}{2}\, {\overline w}\left( {\bf x}_\perp \right) 
+ \kappa\, {\overline n} \left( {\bf x}_\perp \right). 
\label{eqn:initial_energy}
\end{equation}
The distributions ${\overline w}\left( {\bf x}_\perp \right) $ and ${\overline n} \left( {\bf x}_\perp \right)$ are calculated for a given centrality class from the Glauber model. Following the PHOBOS studies of the centrality dependence of the hadron production \cite{Back:2004dy} one uses frequently the value \mbox{$\kappa=0.14$} \cite{Chojnacki:2007rq}. 

By the initial entropy or energy density we understand the values of these quantities at a certain early proper time $\tau=\tau_{\rm i} > 0$  (the time $\tau=0$ corresponds to the moment where the two Lorentz contracted nuclei pass through each other). In the standard hydrodynamic approaches $\sigma_{3\rm i} ({\bf x}_\perp)$ and $\varepsilon_{3\rm i} ({\bf x}_\perp)$ characterize the local equilibrium properties of the system  and $\tau=\tau_{\rm i}$ is treated as the thermalization time, i.e., the time needed to establish local thermodynamic equilibrium. The time evolution for $\tau \geq \tau_{\rm i}$ is fully determined by the hydrodynamic equations supplemented with the appropriate equation of state.

Contrary to standard approaches, in this paper we assume that  the system is not completely thermalized at $\tau=\tau_{\rm i}$. We assume that  only transverse degrees of freedom are thermalized at that time, while the longitudinal motion is described by free streaming. In this case, we may use the following relations connecting the 3D and 2D densities
\begin{equation}
\sigma_{3\rm i}({\bf x}_\perp) = \frac{n_0}{\tau_{\rm i}} \sigma_{2\rm i}({\bf x}_\perp), \quad
\varepsilon_{3 \rm i}({\bf x}_\perp) = \frac{n_0}{\tau_{\rm i}} \varepsilon_{2\rm i}({\bf x}_\perp),
\label{2d3d}
\end{equation}
where  $n_0$ is the normalization constant whose physical interpretation will be given below. Equations (\ref{eqn:initial_entropy}), (\ref{eqn:initial_energy}), and (\ref{2d3d}) suggest the following form of the initial conditions which may be used in our case:
\begin{eqnarray}
\sigma_{2\rm i}({\bf x}_\perp) \;\propto\;  
\frac{1-\kappa}{2}\, {\overline w}\left( {\bf x}_\perp \right) 
+ \kappa\, {\overline n} \left( {\bf x}_\perp \right), \label{eqn:initial_2des} \\
 \varepsilon_{2\rm i} ({\bf x}_\perp) \;\propto\;  
\frac{1-\kappa}{2}\, {\overline w}\left( {\bf x}_\perp \right) 
+ \kappa\, {\overline n} \left( {\bf x}_\perp \right). 
\label{eqn:initial_2den}
\end{eqnarray}
The initial transverse flow at $\tau = \tau_{\rm i}$ is set equal to zero. The dynamics of the system in the proper time interval $\tau_{\rm i} \leq \tau \leq \tau_{\rm tr}$ is determined by the equations of transverse hydrodynamics. Only at $\tau = \tau_{\rm tr}$ a transition to the fully equilibrated phase takes place.

\subsection{Apparent transverse thermalization}
\label{sect:app}

Before we discuss the concept of transverse hydrodynamics, it is useful to recall the arguments for purely transverse thermalization of the systems produced in hadronic and nuclear collisions. 

In the string models, the production of quarks and gluons is often  understood as the result of the Schwinger tunneling mechanism which leads to the Gaussian transverse-momentum distributions. However, as shown by Bialas, the fluctuations of the string tension change the Gaussian dependence into the exponential one \cite{Bialas:1999zg}. The exponential transverse-momentum distributions resemble the Boltzmann distributions and the produced systems may be interpreted as ``transversally'' thermalized. Of course, in this case the thermal character is not a consequence of multiple scattering of the produced partons but the result of the specific mechanism of particle creation (in this context we may talk about {\it apparent} thermalization). 

Analogous features appear also naturally in the color-flux-tube models, where the color fields oscillate \cite{Bialas:1987en,Florkowski:2003mm}. Moreover, in this case the tunneling particles have no longitudinal momentum (in the pair rest frames) \cite{Bialas:1987en}, which implies that the initial ``longitudinal'' pressure is zero. 

A similar physical picture is present in the theory of Color Glass Condensate \cite{Kovchegov:2009he}. In this case, for very early proper time $\tau \ll 1/Q_s$, where $Q_s$ is the saturation scale, the classical gluon fields lead to the following energy-momentum tensor  \cite{Lappi:2006hq,Fukushima:2007ja}
\begin{equation}\label{emt_early}
  T^{\mu\nu} \bigg|_{\tau \ll 1/Q_s} \, = \,
  \left( \begin{array}{cccc} \varepsilon_3 (\tau) & 0 & 0 & 0 \\
      0 &  \varepsilon_3(\tau) & 0 & 0 \\
      0 & 0 & \varepsilon_3 (\tau) & 0  \\
      0 & 0 & 0 & - \varepsilon_3 (\tau) \end{array} \right) \, .
\end{equation}
At later proper times, $\tau \gg 1/Q_s$, both the analytical perturbative approaches \cite{Kovchegov:2005ss} and the full numerical simulations \cite{Krasnitz:2002mn} lead to the form
\begin{equation}\label{emt_late}
  T^{\mu\nu} \bigg|_{\tau \gg 1/Q_s} \, = \,
  \left( \begin{array}{cccc} \varepsilon_3 (\tau) & 0 & 0 & 0 \\
      0 &  \varepsilon_3(\tau)/2 & 0 & 0 \\
      0 & 0 & \varepsilon_3 (\tau)/2 & 0  \\
      0 & 0 & 0 & 0 \end{array} \right) \, .
\end{equation}

The form of the energy-momentum tensor (\ref{emt_late}) and the physics of the string models suggest that during the early evolution of matter the longitudinal pressure is significantly lower that the transverse pressure. These observations trigerred  the development of the concept of the transverse hydrodynamics \cite{Bialas:2007gn,Chojnacki:2007fi}. We note that our framework of the transverse hydrodynamics differs from its first formulation introduced by Heinz and Wong \cite{Heinz:2002rs,Heinz:2002xf}. These differences are discussed in greater detail in \cite{Bialas:2007gn}.

\section{Transverse hydrodynamics}
\label{sect:tr-hyd}

The equations of the transverse hydrodynamics follow from the energy-momentum conservation law, 
\begin{equation}
\partial_\mu T_2^{\mu \nu}=0,
\label{hyd2d}
\end{equation}
with the energy-momentum tensor defined by the formula \cite{Ryblewski:2008fx}
\begin{equation}
T_2^{\mu \nu} = \frac{n_0}{\tau} \left[
\left(\varepsilon _2 + P_2\right) U^{\mu}U^{\nu} 
- P_2 \,\,\left( g^{\mu\nu} + V^{\mu}V^{\nu} \right)\,\, \right].
\label{tensorT1}
\end{equation}
The normalization constant $n_0$ may be interpreted as the density of transverse clusters in rapidity. The clusters are formed by groups of partons having the same rapidity. They are 2D objects, whose thermodynamic properties are described by the 2D thermodynamic variables: $\varepsilon_2$, $P_2$, $\sigma_2$ and $T_2$ (2D energy density, pressure, entropy density, and temperature, respectively). These quantities  satisfy the standard thermodynamic identities:
\begin{eqnarray}
\varepsilon_2 + P_2 &=& T_2 \,\sigma_2, \nonumber \\
d\varepsilon_2 &=& T_2 \,d\sigma_2,  \nonumber \\
dP_2 &=& \sigma_2 \,dT_2.
\label{2dthermo}
\end{eqnarray}
The baryon chemical potential is neglected here, since we consider the midrapidity region where the baryon density is very small. The definition of the energy-momentum tensor (\ref{tensorT1}) contains the two four-vectors,
\begin{eqnarray}
U^{\mu} &=& ( u_0 \cosh\eta,u_x,u_y, u_0 \sinh\eta), \nonumber \\
V^{\mu} &=& (\sinh\eta,0,0,\cosh\eta),
\label{U}
\end{eqnarray}
where 
\begin{equation}
u^\mu = \left(u^0, {\vec u}_\perp, 0 \right) = \left(u^0, u_x, u_y, 0 \right)
\end{equation}
is the hydrodynamic flow in the plane $z=0$, while $\eta$ is the spacetime rapidity defined by Eq.~(\ref{yandeta}). The four-vectors $U^\mu$ and $V^\mu$ are normalized in the following way
\begin{eqnarray}
U^\mu U_\mu &=& 1, \quad V^\mu V_\mu = -1, \quad U^\mu V_\mu = 0.
\label{UV}
\end{eqnarray}
The four-vector $U^\mu$ combines the motion of the fluid element in a cluster with the motion of the cluster, thus it corresponds to the flow four-velocity in the standard hydrodynamics.  The term $V^\mu V^\nu$ in (\ref{tensorT1}) is responsible for vanishing of the longitudinal pressure, i.e., in the local rest-frame of the fluid element, where we have $U^\mu = (1,0,0,0)$ and $V^\mu = (0,0,0,1)$, one finds
\begin{equation}
T^{\mu \nu}_2 = \frac{n_0}{\tau} \left(
\begin{array}{cccc}
\varepsilon _2 & 0 & 0 & 0 \\
0 & P_2 & 0 & 0 \\
0 & 0 & P_2 & 0 \\
0 & 0 & 0 & 0
\end{array} \right).
\label{tensorT2}
\end{equation}
One can notice that exactly the same structure of the energy-momentum tensor appears in the theory of the {\it color glass condensate} and {\it glasma} for \mbox{$\tau \gg 1/Q_s$}, see Eq.~(\ref{emt_late})

We solve the equations of the transverse hydrodynamics numerically using the equation of state~\footnote{The form of the equation of state used in this paper is valid for bosons. For the classical massless  particles on the plane \mbox{$P_2 = \nu_g T_2^3/(2 \pi)$}, and for the massless fermions \mbox{$P_2 = 3 \nu_g T_2^3 \zeta(3)/(8 \pi)$} \cite{Ryblewski:2008fx}. }
\begin{equation}
P_2 = \frac{\varepsilon_2}{2}  = \frac{\nu_g \zeta(3)}{2\pi} T_2^3 ,
\label{eos2d}
\end{equation}
where $\nu_g = 16$ reflects the spin and color degeneracy of gluons. Our investigations are restricted to the midrapidity region ($z \approx \eta \approx 0$), where the partonic system may be treated as boost-invariant.  

The structure of the energy-momentum tensor (\ref{tensorT1}) suggests that it is convenient to introduce the three-dimensional densities of the transversally thermalized system,
\begin{equation}
\varepsilon_3^{\rm tr} = \frac{n_0}{\tau} \varepsilon_2, \quad 
\sigma_3^{\rm tr} = \frac{n_0}{\tau} \sigma_2, \,\,\, \ldots \,\,\, .
\label{32therm}
\end{equation}
With the help of this notation, the equations of transverse hydrodynamics take the form
\begin{eqnarray}
U^\mu \partial_\mu \left( T_2 U^\nu \right) &=& \partial^\nu T_2 + V^\nu V^\mu \partial_\mu T_2, \nonumber \\
\partial_\mu \left( \sigma_3^{\rm tr} U^\mu \right) &=& 0.
\label{trhydeq}
\end{eqnarray}
Here the upper equation is the analog of the Euler equation in classical hydrodynamics, while the lower equation describes the conservation of the entropy. Using the equation of state (\ref{eos2d}) we may rewrite the entropy-conservation law in the form
\begin{equation}
\partial_\mu \left( \frac{n_0}{\tau} T_2^2 U^\mu \right) = 0.
\label{trhydeqS}
\end{equation} 
The structure of Eqs. (\ref{trhydeq}) and (\ref{trhydeqS}) indicates that they are scale invariant --- the flow profile does not change if the temperature is multiplied by an arbitrary constant. We shall use this property later to match properly the transverse hydrodynamics with the standard perfect-fluid hydrodynamics.

\section{Landau matching conditions}
\label{sect:Landau-match}

The large collection of the RHIC data (transverse-momentum spectra, ratios of hadronic abundances, etc.) suggests that the system produced finally in Au+Au collisions at the highest beam energy of 200 GeV per nucleon pair is very well thermalized. For our modeling this means that, at a certain stage, the initial transversally thermalized system must undergo the full 3D equilibration. In a microscopic approach we expect that the 3D equilibration is a gradual process. In our effective model, however, we treat this transition as a sudden change from the transverse to the standard perfect-fluid hydrodynamics. We stress that this approach is an {\it approximation where the continuous process is replaced by a delayed step-like change}. 

The sudden equilibration transition is described by the Landau matching condition
\begin{equation}
T_2^{\mu \nu} U_\nu = T^{\mu \nu}_{3} U_\nu, \label{LMc1}
\end{equation} 
where $T^{\mu \nu}_{3}$ is the standard energy-momentum tensor of the perfect-fluid hydrodynamics~\footnote{The use of the same four-vector $U^\mu$ in the definitions of $T_2^{\mu \nu}$ and $T_3^{\mu \nu}$ ensures that the energy flux on both sides of the transition surface has the same direction.   }
\begin{equation}
T_3^{\mu \nu} = (\varepsilon_3 + P_3) U^\mu U^\nu - P_3 g^{\mu \nu}. \label{LMc2}
\end{equation}  
Here $\varepsilon_3$ and $P_3$ are the three-dimensional (3D) energy density and pressure of the system immediately after the equilibration transition. We assume that they are given by the equation of state constructed in Ref. \cite{Chojnacki:2007jc}.

Equations (\ref{LMc1}) and (\ref{LMc2}) give
\begin{equation}
\varepsilon_3^{\rm tr} = \frac{n_0}{\tau} \varepsilon_2 = \varepsilon_3, \label{LMc3}
\end{equation} 
which should be supplemented by the requirement of the entropy growth,
\begin{equation}
\sigma_3^{\rm tr} = \frac{n_0}{\tau} \sigma_2 \leq \sigma_3, \label{LMc4}
\end{equation} 	
where $s_3$ is the 3D entropy density. Dividing both sides of Eqs. (\ref{LMc3}) and (\ref{LMc4}) one obtains
\begin{equation}
T_2 \geq \frac{3 \,\varepsilon_3}{2 \,\sigma_3}. \label{LMc5}
\end{equation} 	

Certainly, our treatment of the full equilibration transition is very much simplified. More elaborate approaches would describe this kind of transformation using kinetic theory or dissipative hydrodynamics \cite{Kovchegov:2005az,Bozek:2007di,Zhang:2008kj}. For example, in the dissipative hydrodynamics the viscosity corrections increase transverse pressure and decrease longitudinal pressure. In the boost-invariant one-dimensional case we find \cite{Kovchegov:2005az}
\begin{equation}\label{emt_visc}
  T^{\mu\nu} \, = \,
  \left( \begin{array}{cccc} \varepsilon_3 & 0 & 0 & 0 \\
      0 &  P_3 + \frac{2}{3} \frac{\eta}{\tau} & 0 & 0 \\
      0 & 0 & P_3 + \frac{2}{3} \frac{\eta}{\tau} & 0  \\
      0 & 0 & 0 & P_3 - \frac{4}{3} \frac{\eta}{\tau} \end{array} \right) \, ,
\end{equation}
where $\eta$ is the shear viscosity. The matching of the tensor (\ref{emt_visc}) with the forms (\ref{emt_early}) or (\ref{emt_late}) requires, however, very large values of $\eta$. This observation questions strict applicability of the dissipative hydrodynamics at the very early stages. Thus, we are of the opinion that our approach based on the use of the transverse hydrodynamics offers an interesting alternative for viscous hydrodynamics in this case. 

As shown below, the important aspect of our procedure is that the sudden equilibration transition may take place at $\tau_{\rm tr}$ of about 1 fm. In our opinion, this result indicates that the ``equivalent'' gradual equilibration processes may be extended in time and no assumption about the sudden and full thermalization of the system at very short times ($\tau_{\rm i}$ of about 0.2 fm for perfect-fluid hydrodynamics) must be made.

\section{Perfect-fluid hydrodynamics and freeze-out}
\label{sect:perfect-fluid}

We assume that after the transition time $\tau_{\rm tr}$ the matter behaves like a perfect fluid. The spacetime evolution is described with the help of the formalism developed in Ref.~\cite{Chojnacki:2006tv}. Its characteristic feature is the use of the modern equation of state which interpolates between the lattice QCD results and the hadron-gas calculations \cite{Chojnacki:2007jc}. We note that a similar equation of state has been constructed very recently in Ref. \cite{Huovinen:2009yb}.

In a certain way, we may treat the Landau matching conditions as a method to deliver initial conditions for the perfect-fluid hydrodynamics at $\tau = \tau_{\rm tr}$. Since the system has spent already some time in the expanding ``transverse phase'', the transverse flow at $\tau = \tau_{\rm tr}$ is not zero (the issue of the initial non-zero transverse flow is discussed in \cite{Chojnacki:2004ec,Sinyukov:2006dw,Gyulassy:2007zz}). In fact, this flow is relatively large because the development of the transverse flow in the transverse hydrodynamics is fast (larger sound velocity, no losses of energy due to the longitudinal work, etc.). Of course, the comparison with the data eventually constraints the amount of the transverse flow that is allowed to be generated in the transversally thermalized phase. 

The physical observables are obtained by using the Cooper-Frye formula with the hypersurface defined by the constant value of the freeze-out temperature, \mbox{$T_{\rm 3 f}$ = const}. Thus, we adopt the single freeze-out scenario and do not distinguish between the chemical and thermal freeze-outs \cite{Broniowski:2001we}. This is of course an approximation that is known to work well for pions and worse for protons \cite{Nonaka:2006yn,Broniowski:2008vp}.  We note that only the fully equilibrated region of the spacetime is used to construct the freeze-out hypersurfce (in other words, there is no hadronic emission from the transversally thermalized stage for $\tau < \tau_{\rm tr}$).

The thermal Monte-Carlo code {\tt THERMINATOR} \cite{Kisiel:2005hn} is used to generate primordial particles that include stable hadrons and all known hadronic resonances. {\tt THERMINATOR} simulates the decays of resonances which proceed in cascades. All studied observables (spectra, the elliptic flow $v_2$, and the HBT radii) are calculated with the help of the Monte-Carlo method. In particular, the femtoscopic observables are obtained with the help of the two-particle Monte-Carlo method, where the correlations between the pions are introduced by the appropriate weights defined by the squares of the two-particle wave functions \cite{Kisiel:2006is}. 

\section{Restricting the model parameters}
\label{sect:restricting}

Our model has altogether nine parameters. The main six parameters are: $n_0$ -- the overall normalization, $T_{2\rm i}$ -- the initial central temperature of the transversally thermalized system, $T_{3\rm f}$ -- the freeze-out temperature,  $\tau_{\rm i}$ -- the initial proper time, $\tau_{\rm tr}$ -- the 2D $\to$ 3D transition time when the Landau matching conditions are applied, and $\kappa$ -- the parameter describing the admixture of the binary-collision density in the initial conditions (\ref{eqn:initial_2des}) and (\ref{eqn:initial_2den}). Similarly to Ref.~\cite{Broniowski:2008qk}, we use the freeze-out temperature
\begin{equation}
T_{3\rm f} = 140 \, \hbox{MeV}.
\label{T3f}
\end{equation}
In the thermal generation of the particles on the freeze-out hypersurface, the extra three chemical potentials are also introduced: $\mu_B$ -- the baryon chemical potential, $\mu_S$ -- the strangeness chemical potential, and $\mu_{I_3}$ -- the isospin chemical potential. The values of the chemical potentials are much smaller than $T_{3\rm f}$, hence, their effect on the evolution of matter is neglected. They appear only in the thermal distribution functions used to generate particles on the freeze-out hypersurface. Following Refs. \cite{Florkowski:2001fp} we use the values
\begin{equation}
\mu_B = 28.5 \,\hbox{MeV}, \, \mu_S = 6.9 \,\hbox{MeV}, \, \mu_{I_3} = -0.9 \,\hbox{MeV}.
\label{fout-chem}
\end{equation}

Our numerical calculations were performed for the initial conditions of the form (\ref{eqn:initial_2des}) and (\ref{eqn:initial_2den}). A better agreement with the data (especially for the elliptic flow) was obtained generally for the case (\ref{eqn:initial_2den}), hence, in the following we shall restrict our considerations to the situation where
\begin{eqnarray}
\varepsilon_2\left(\tau_{\rm i},{\vec x}_\perp \right) 
&=& \frac{\nu_g \,  \zeta(3) }{\pi} \, T_{2}^{\,3} \left(\tau_{\rm i},{\vec x}_\perp \right) \nonumber \\
&&\, \propto \,  \frac{1-\kappa}{2} {\overline w}\left( {\bf x}_\perp \right)  + \kappa {\overline n}\left( {\bf x}_\perp \right) .
\label{initcond}
\end{eqnarray}	
The normalization constant required in (\ref{initcond}) determines the 2D initial central temperature of the system, \mbox{$T_{2 \,\rm i}=T_2(\tau_{\rm i},0)$}. 

The selection of the optimal values of \mbox{$T_{2 \,\rm i}$} and $\kappa$ in (\ref{initcond}) will be discussed in more detail below. We want to stress now the fact that as long as we ignore the Landau matching condition for the entropy our results are insensitive with respect to the following rescaling of the parameters $n_0$ and $T_{2 \,\rm i}$,
\begin{equation}
n_0 \to \lambda n_0, \quad T_{2 \,\rm i} \to \lambda^{-1/3} T_{2 \,\rm i}.
\label{scale1}
\end{equation}
\begin{widetext}

\begin{figure}[t]
\begin{center}
\subfigure{\includegraphics[angle=0,width=0.33\textwidth]{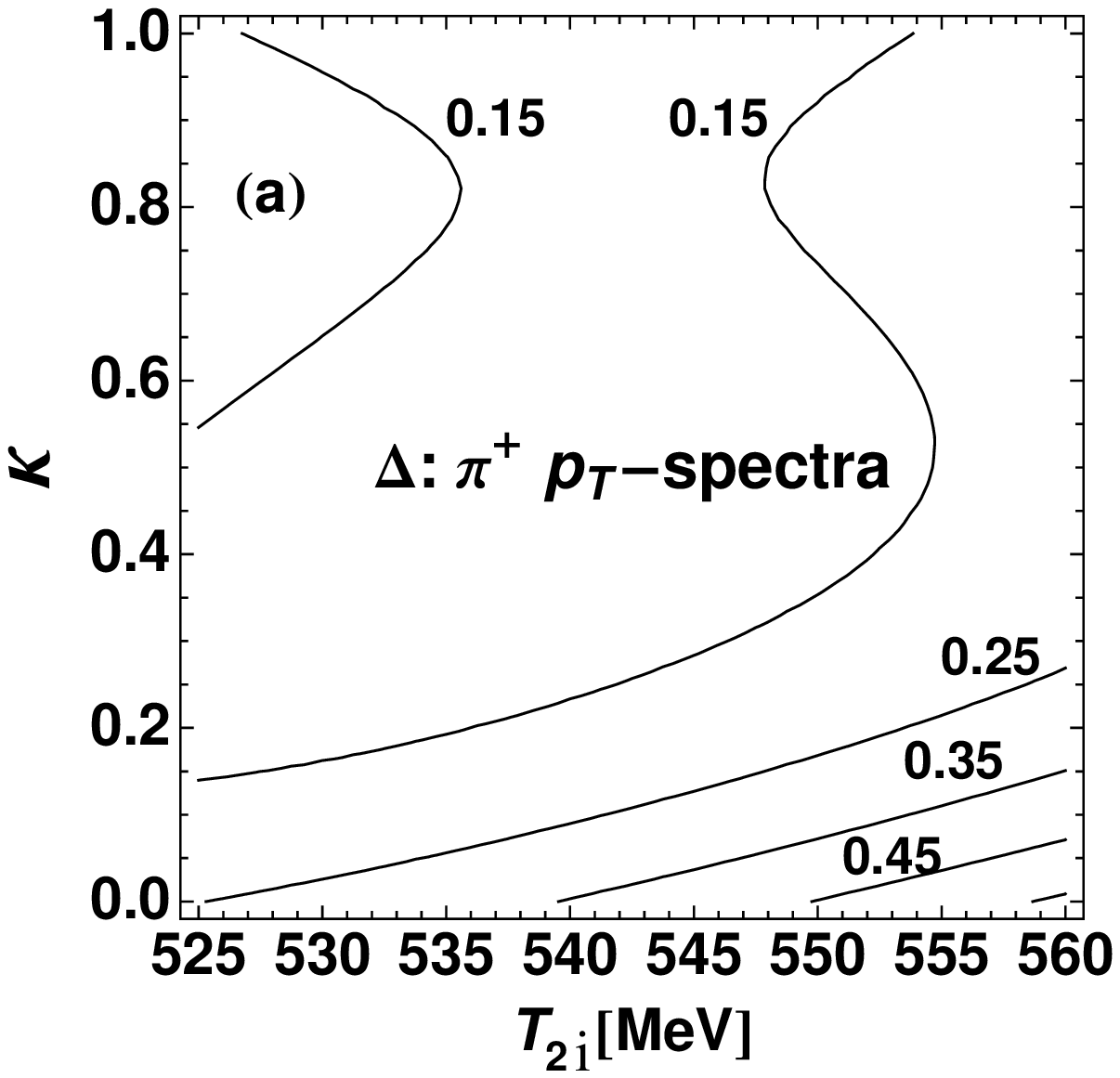}} \hspace{1.0cm}
\subfigure{\includegraphics[angle=0,width=0.33\textwidth]{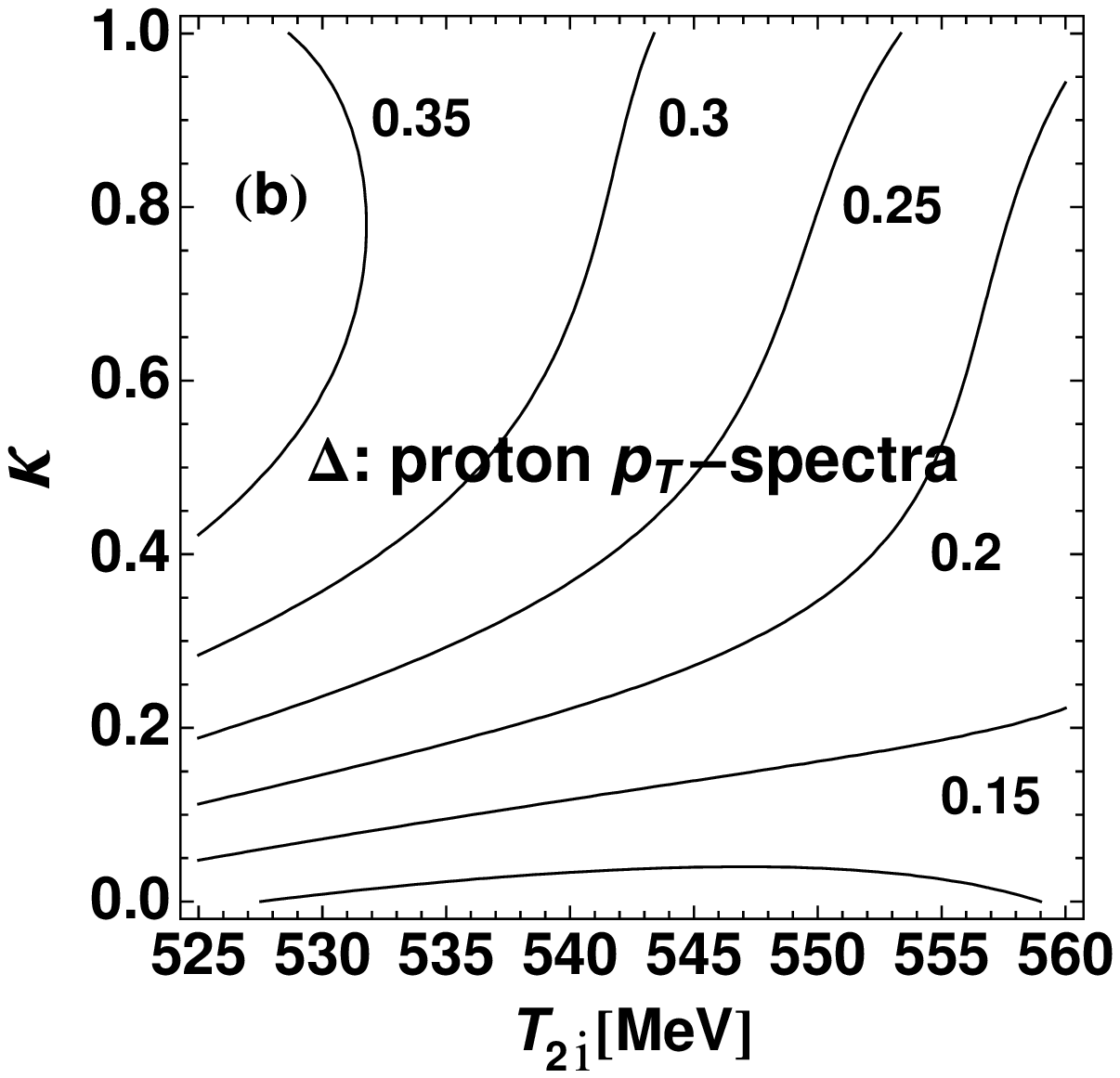}} \\
\subfigure{\includegraphics[angle=0,width=0.33\textwidth]{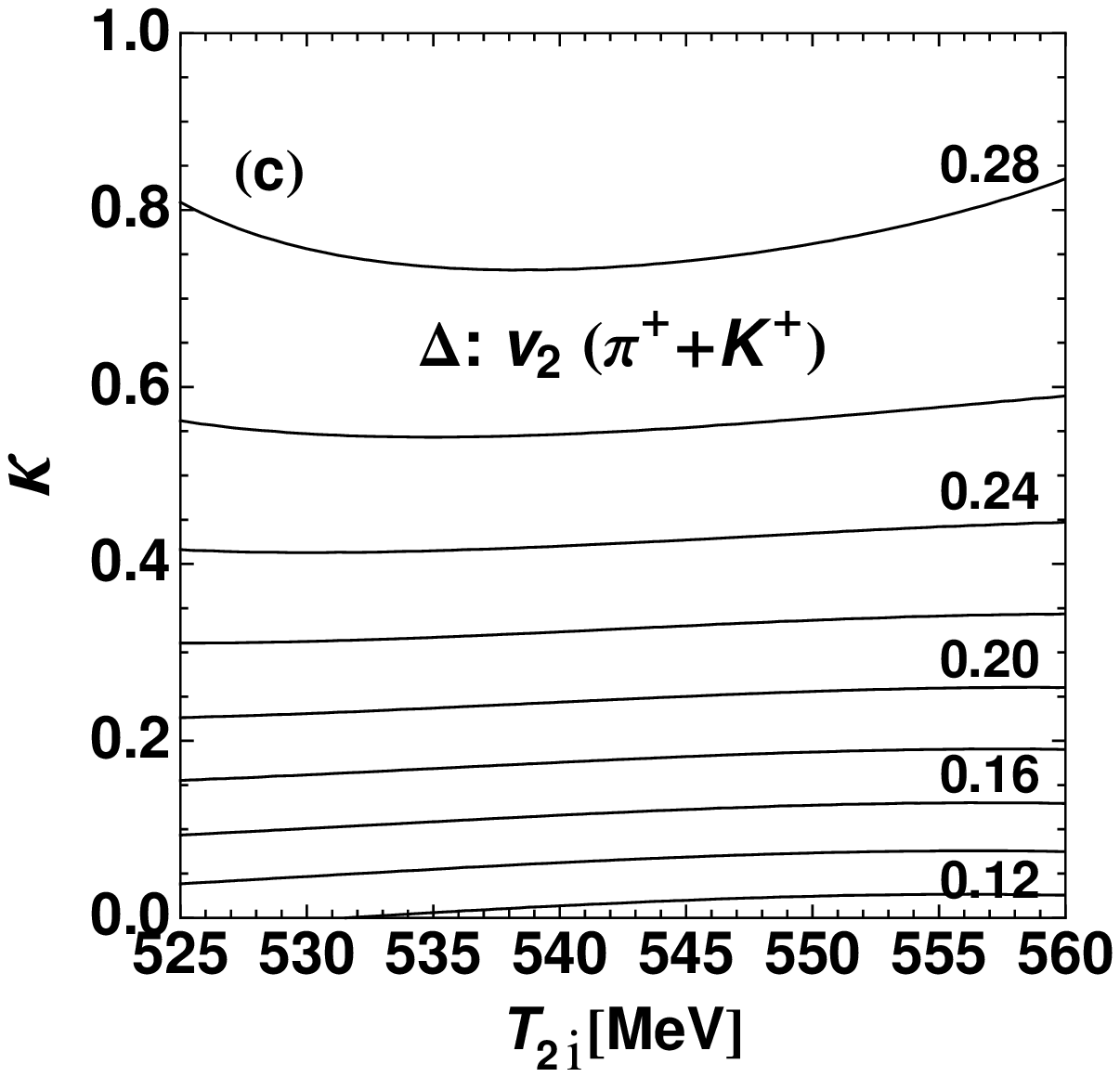}} \hspace{1.0cm}
\subfigure{\includegraphics[angle=0,width=0.33\textwidth]{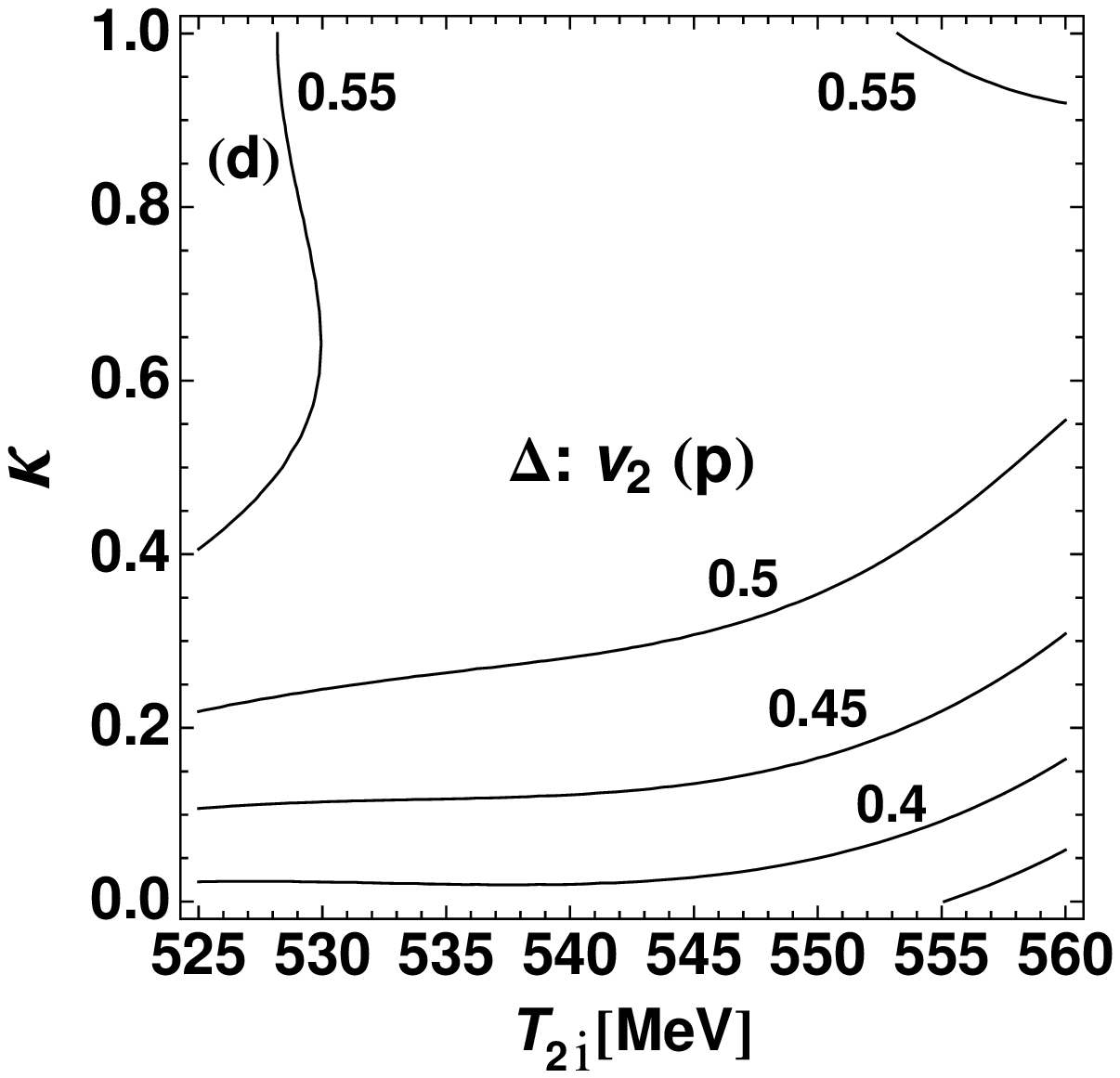}} \\
\subfigure{\includegraphics[angle=0,width=0.33\textwidth]{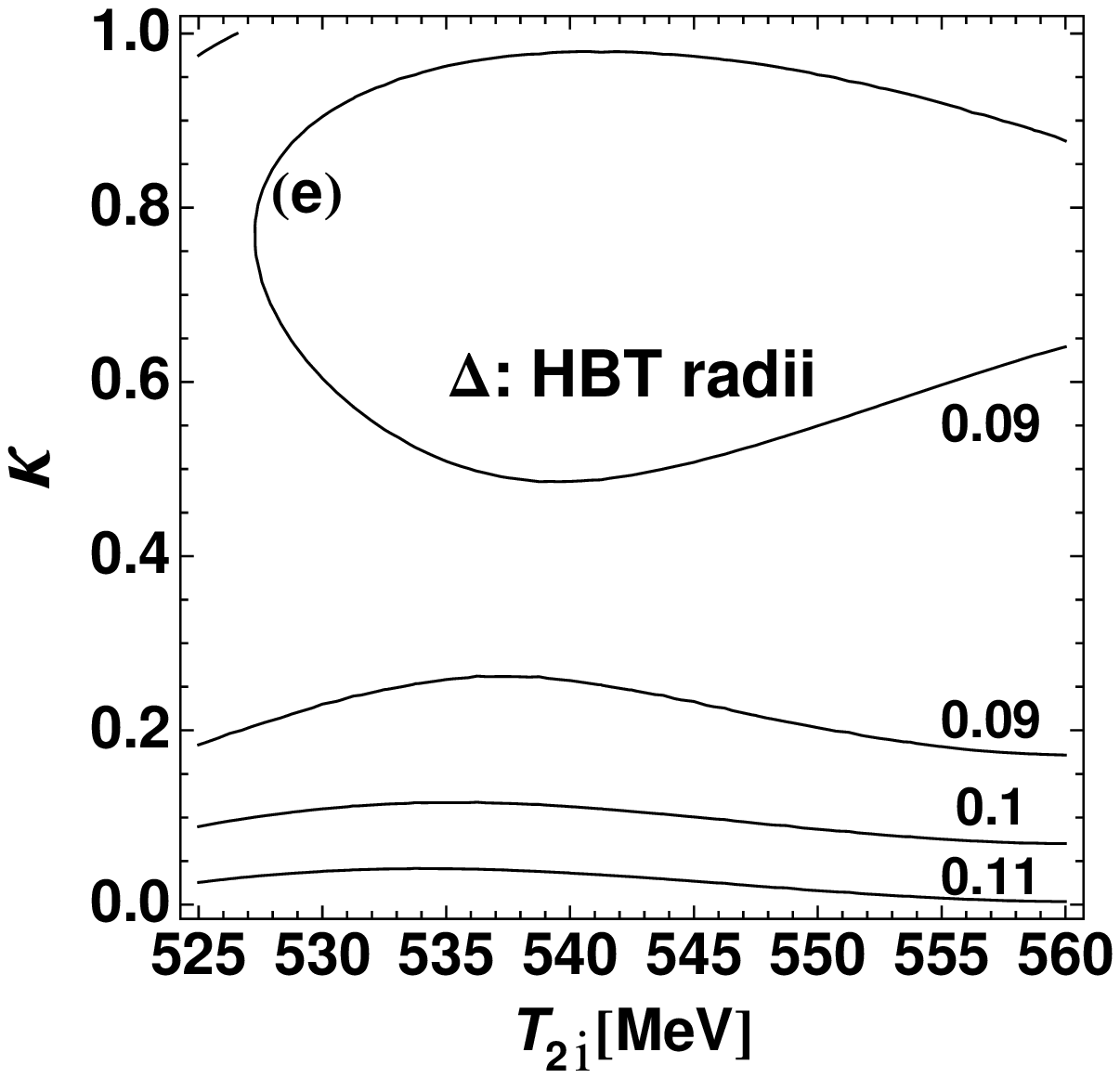}} \hspace{1.0cm}
\subfigure{\includegraphics[angle=0,width=0.33\textwidth]{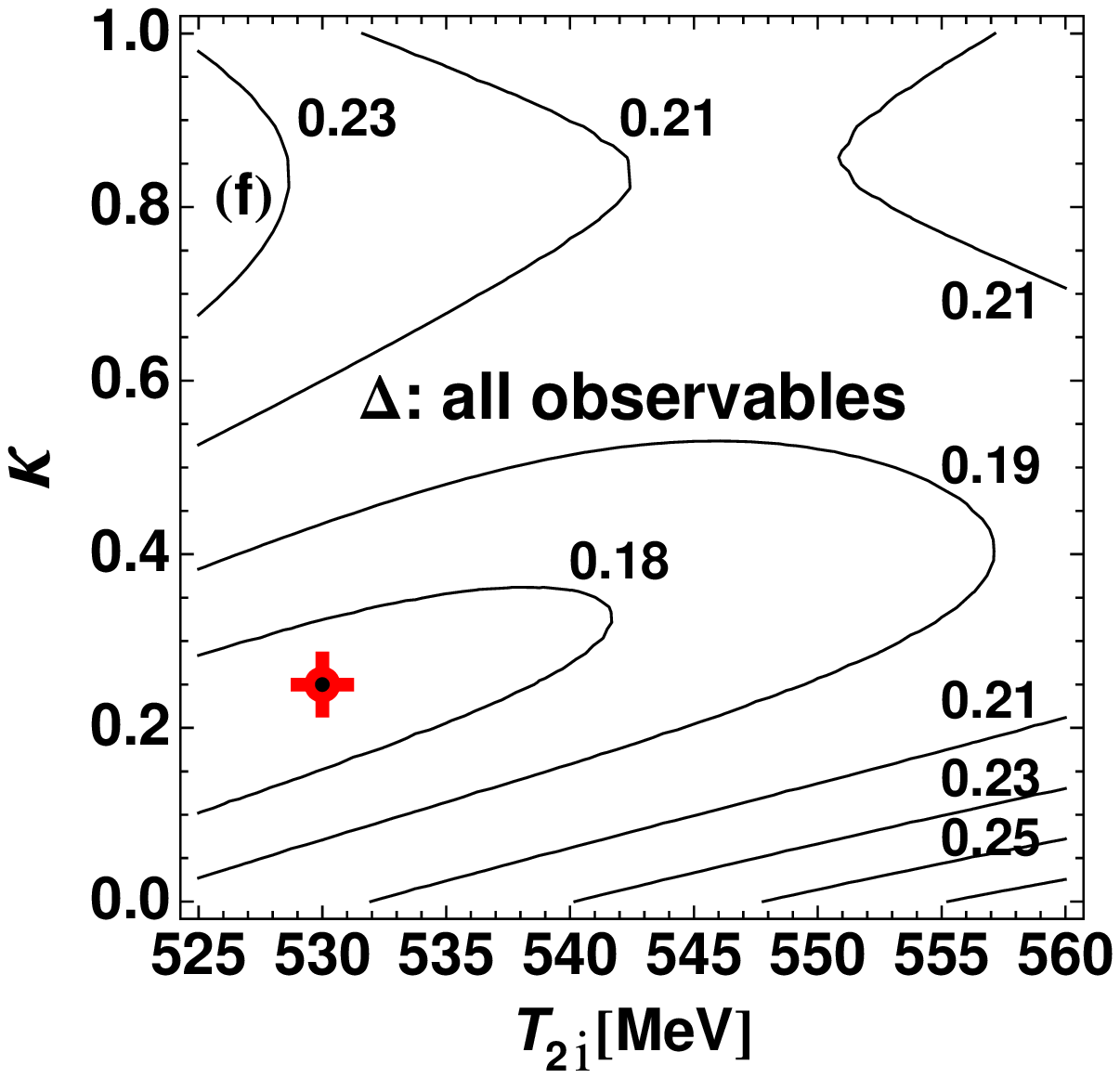}}
\end{center}
\caption{(Color online) Contour plots of the quantities $\Delta$ describing the average relative differences between the model results and the data, see Eqs.~(\ref{Delta}) and (\ref{Delta12}). The values of $\Delta$ are shown as functions of $\kappa$ and $T_{2\rm i}$. The two upper plots show the agreement for the pion and proton spectra, while the two middle plots show the agreement for the pion+kaon and proton $v_2$. The lower plot on the left-hand-side shows the agreement for the HBT radii, and the lower plot on the right-hand-side shows the agreement for the all considered observables (spectra, $v_2$, and the HBT radii). The calculations were done with the parameters specified by Eqs. (\ref{T3f}), (\ref{fout-chem}), (\ref{n0}), and (\ref{time-par}) for the centrality class $c = 20-30\%$ (the impact parameter $b$ = 7.16 fm). The data used to calculate $\Delta$ were taken from Refs.~\cite{Adler:2003cb,Adler:2003kt,Adams:2004yc}. }
\label{fig:contours}
\end{figure}

\end{widetext}
This transformation does not change the energy density and flow of the transversally thermalized system. On the other hand, it affects the entropy since we have
\begin{equation}
\frac{n_0}{\tau} \sigma_2 \to \lambda^{1/3} \frac{n_0}{\tau} \sigma_2.
\end{equation}
Thus, if a satisfactory description of the data is achieved with certain values of $n_0$ and $T_{2 \,\rm i}$ (with the conserved energy during the 2D $\to$ 3D transition) we may always use the transformation (\ref{scale1}) to change the entropy in the purely transverse stage in such a way that it grows during the equilibration transition.  By decreasing $\lambda$  we effectively reduce the number of clusters and make them hotter. The decays of such ``more organized'' and hotter clusters produce more entropy.

Following our previous studies \cite{Ryblewski:2009hm,Florkowski:2009wb} where the transverse hydrodynamics was followed by the sudden equilibration and freeze-out (without the perfect-fluid stage) we used the value
\begin{equation}
n_0 = 0.43,
\label{n0}
\end{equation}
which turned out to be satisfactory also in the present studies. 

We have also found that the optimal values of the parameters $\tau_{\rm i}$ and $\tau_{\rm tr}$ are
\begin{equation}
\tau_{\rm i} = 0.25 \,\hbox{fm}, \quad \tau_{\rm tr} = 1.0 \,\hbox{fm}.
\label{time-par}
\end{equation}
The evolution times of the transversally thermalized system that are longer than 0.75 fm  lead to too strong radial flow. On the other hand, a slightly better description of the data is achieved if the transverse-hydrodynamics stage is implemented earlier. Such constraints lead to the choice (\ref{time-par}). We note that the time interval of the purely transverse hydrodynamics considered in this paper is the same as the time for the initial free streaming considered in Ref.~\cite{Broniowski:2008qk}. 

The quality of the fits discussed in this Section is characterized by the parameters $\Delta$ defined as the average relative difference between the model results and the data
\begin{equation}
\Delta = \frac{1}{N} \sum_{i=1}^N \frac{|x^{\rm th}_i - x^{\rm exp}_i|}{x^{\rm exp}_i}.
\label{Delta}
\end{equation} 
Here $x^{\rm exp}_i$ is the $i$th experimental point (e.g., the value of the transverse-momentum spectrum or the value of the elliptic flow) and $x^{\rm th}_i$ is the corresponding theoretical value. The quantity $N$ is the number of the experimental points (e.g., the number of the points in the measured transverse-momentum spectrum). The values of $\Delta$ are calculated for different observables separately, and also for combined observables. In the latter case we define $\Delta$ as the average
\begin{equation}
\Delta_{\rm obs1+obs2} = \frac{1}{2} \left( \Delta_{\rm obs1}+\Delta_{\rm obs2}\right).
\label{Delta12}
\end{equation}
We stress that the lack of knowledge about the systematic errors (in the case of the transverse-momentum data presented in \cite{Adler:2003cb}) or discrapancies between the systematic errors given by different experiments (e.g., the HBT results given in \cite{Adams:2004yc} and \cite{Adler:2004rq} makes a thorough $\chi^2$ analysis quite difficult. It requires making comparisons between different experiments and such a study goes beyond the problems discussed in the present paper.

\subsection{Determination of \mbox{$T_{2 \,\rm i}$} and $\kappa$}

In order to illustrate the choice of the optimal values of \mbox{$T_{2 \,\rm i}$} and $\kappa$, in Fig.~\ref{fig:contours} we show the values of $\Delta$ calculated for different physical observables. As discussed above, the values of other parameters were motivated by our earlier hydrodynamic studies and are given by Eqs.~(\ref{T3f}), (\ref{fout-chem}), (\ref{n0}), and (\ref{time-par}) \footnote{The multidimensional fits using all parameters are very much time consuming and not possible  for us. In particular, the calculations of the HBT radii are very long. } The model calculations are compared with the data published in Refs.~\cite{Adler:2003cb,Adler:2003kt,Adams:2004yc}. We consider the centrality class $c = 20-30\%$ (the impact parameter $b$ = 7.16 fm)

\begin{figure}[t]
\begin{center}
\includegraphics[angle=0,width=0.5 \textwidth]{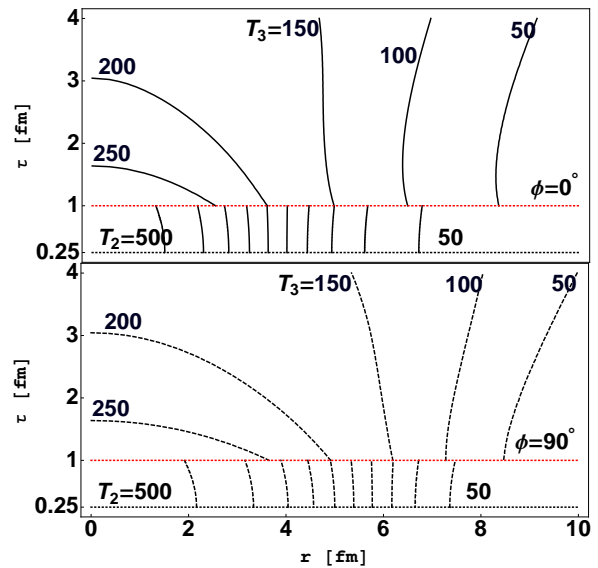}
\end{center}
\vspace{-3mm}
\caption{(Color online) The 2D and 3D isotherms for $\phi=0$ (upper part) and $\phi=\pi/2$ (lower part). The values of the parameters are specified by Eqs. (\ref{T3f}), (\ref{fout-chem}), (\ref{n0}), (\ref{time-par}), and (\ref{T2i-kappa}). The centrality class $c = 20-30\%$.}
\label{fig:isotherms}
\end{figure}
\begin{figure}[t]
\begin{center}
\includegraphics[angle=0,width=0.4 \textwidth]{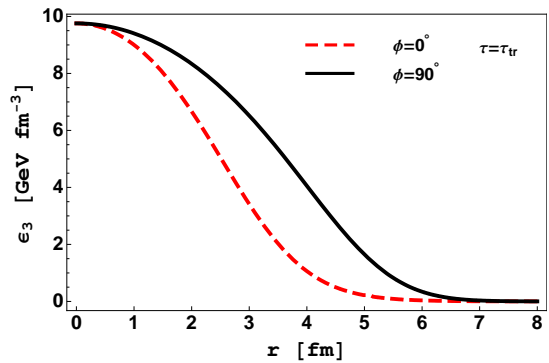}
\end{center}
\vspace{3mm}
\caption{(Color online)  The energy density at the transition time $\tau_{\rm tr}$ plotted as a function of $r$ for $\phi = 0$ (dashed line) and $\phi = 90^0$ (solid line). The parameters the same as in Fig. \ref{fig:isotherms}. }
\label{fig:energy}
\end{figure}

The upper two plots of Fig.~\ref{fig:contours} show the quality of fitting the transverse momentum spectra. The experimental PHENIX results used to calculate $\Delta$ are taken from Ref. \cite{Adler:2003cb}. We find that the pion spectra are described with the accuracy of about 10--60\% depending on the values of \mbox{$T_{2 \,\rm i}$} and $\kappa$. A similar accuracy may be achieved in the fits of the proton spectra, where \mbox{$\Delta$ = 10--40\%}. The interesting point is, however, that the pion and proton spectra favor different values of \mbox{$T_{2 \,\rm i}$} and $\kappa$. It turns out that it is impossible to achieve the accuracy better than 20\% for both pions and protons. 

The middle two plots show the quality of fitting the elliptic flow coefficient $v_2$. The PHENIX data are taken now from Ref.~\cite{Adler:2003kt}. In the considered ranges of the parameters \mbox{$T_{2 \,\rm i}$} and $\kappa$ we find that the pion+kaon $v_2$ is reproduced at the level of about 10-30\%, while the proton $v_2$ is reproduced much worse, at the level of about 30-60\%. We note that such high values of the proton elliptic flow are characteristic for the calculations using the crossover phase transition \cite{Chojnacki:2007rq,Huovinen:2009yb}. The difficulties connected with the correct reproduction of the proton $v_2$ are illustrated in the panels (a) and (d) of Fig. \ref{fig:contours}  where one can see that the good fits of the pion spectra and proton $v_2$ are anticorrelated.

\begin{figure}[t]
\begin{center}
\includegraphics[angle=0,width=0.4 \textwidth]{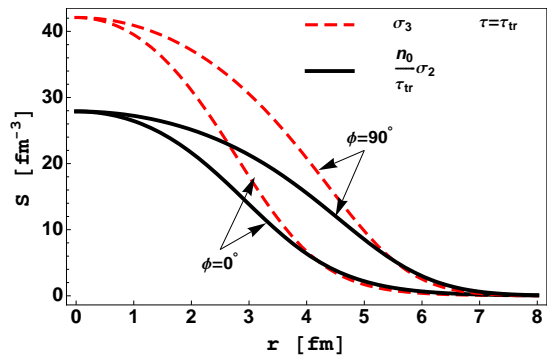}
\end{center}
\vspace{3mm}
\caption{(Color online) The entropy density just before (solid lines) and after (dashed lines) the transition time  $\tau_{\rm tr}$ plotted as a function of $r$ for $\phi = 0$  and $\phi = 90^0$. The parameters the same as in Fig. \ref{fig:isotherms}. }
\label{fig:entropy}
\end{figure}
\begin{figure}[t]
\begin{center}
\includegraphics[angle=0,width=0.4 \textwidth]{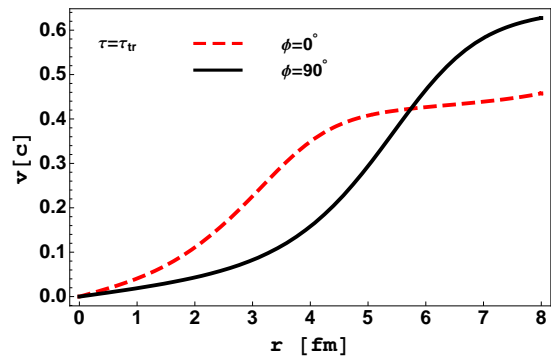}
\end{center}
\vspace{3mm}
\caption{(Color online)  The transverse flow at the transition time $\tau_{\rm tr}$ plotted as a function of $r$ for $\phi = 0$ (dashed line) and $\phi = 90^0$ (solid line). The parameters the same as in Fig. \ref{fig:isotherms}. }
\label{fig:flow}
\end{figure}

In the lower left part of Fig.~\ref{fig:contours} we show the agreement between the model and experimental results for the HBT radii. In this case the STAR data from Ref.~\cite{Adams:2004yc} are used. Generally, we find a very good description, at the level of 10\% for all the values of \mbox{$T_{2 \,\rm i}$} and $\kappa$. The good description of the HBT results is a consequence of many factors \cite{Pratt:2009hu}. In our case they include the use of the semi-hard realistic equation of state \cite{Chojnacki:2007jc} and the implementation of the "transverse" phase that speeds up the time evolution of the system.

The analysis presented in this Section suggests that the optimal values of the parameters \mbox{$T_{2 \,\rm i}$} and $\kappa$ are
\begin{eqnarray}
T_{2 \,\rm i} &=& 530 \, \hbox{MeV}, \nonumber \\
\kappa &=& 0.25.
\label{T2i-kappa}
\end{eqnarray}
These two values together with the earlier specified parameters, see Eqs. (\ref{T3f}), (\ref{fout-chem}), (\ref{n0}), and (\ref{time-par}), will be used in the following presentation.

We note that Fig. \ref{fig:contours} illustrates quantitatively difficulties of obtaining a consistent good fit to all studied observalbles --- different physical quantities are described well by different ranges of the parameters.

\section{Spacetime behavior of 2D and 3D thermodynamic quantities}
\label{sect:spacetime}

Figure~\ref{fig:isotherms} shows an example of the spacetime evolution of matter in the vicinity of the transition time $\tau_{\rm tr}$. The values of the parameters are specified by Eqs. (\ref{T3f}), (\ref{fout-chem}), (\ref{n0}), (\ref{time-par}), and (\ref{T2i-kappa}). We recall that they correspond to the centrality class $c = 20-30\%$ (the impact parameter $b$ = 7.16 fm). In this case the transverse hydrodynamics starts at $\tau_{\rm i}$ = 0.25 fm and continues till $\tau = \tau_{\rm tr}$ = 1 fm, when the transition to 3D evolution takes place. 

The upper and lower parts of Fig. \ref{fig:isotherms} show the 2D and 3D isotherms for $\phi$ = 0 and $\phi=90^\circ$, respectively. The energy density which is the same on both sides of the transition is plotted in Fig. \ref{fig:energy}.

The application of the Landau matching conditions implies that there is no simple connection between the 2D and 3D isotherms at $\tau = \tau_{\rm tr}$. However, if the hadronic gas was described by the equation of state of massless particles then Eq. (\ref{LMc5}) would yield the condition $T_2 > (9/8) T_3$. Thus, we expect qualitatively that $T_2$ should be larger than $T_3$ in the transition region. By inspection of Fig. \ref{fig:isotherms} we see that this condition is fulfilled for the central part of the fireball (for $r \lesssim 5$ fm if $\phi=0$ and for $r \lesssim 6$ fm if $\phi=90^0$).

The exact jump of the entropy density is illustrated in Fig. \ref{fig:entropy}, where the solid lines describe the entropy density just before the 2D $\to$ 3D transition and the dashed lines describe the entropy just after the transition. In agreement with the expectations formulated above, we observe that the entropy increases substantially in the central part of the fireball. In the outer parts of the fireball the entropy densities in the transverse and perfect-fluid phases are very much similar. For large values of $r$ the entropy in the transverse phase becomes smaller and the Landau matching condition is violated. However, this region is excluded from the physical considerations as it corresponds to the 3D temperatures smaller than the freeze-out temperature.  We note that the entropy  in the transverse phase can be made even smaller if the scaling (\ref{scale1}) with $\lambda < 1$ is performed.

For completeness, in Fig. \ref{fig:flow} we present the flow profiles at the transition time. By construction the flow is the same on both sides of the transition surface. 

\begin{figure}[t]
\begin{center}
\includegraphics[angle=0,width=0.45 \textwidth]{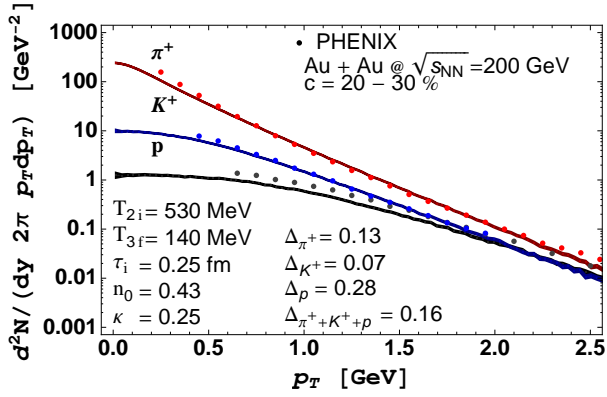}
\end{center}
\vspace{-3mm}
\caption{(Color online) The comparison of the model and experimental transverse-momentum spectra of pions, kaons, and protons. The data are taken from Ref. \cite{Adler:2003cb}. The model parameters are defined by Eqs. (\ref{T3f}), (\ref{fout-chem}), (\ref{n0}), (\ref{time-par}), and (\ref{T2i-kappa}).}
\label{fig:ppt}
\end{figure}
\begin{figure}[t]
\begin{center}
\includegraphics[angle=0,width=0.45 \textwidth]{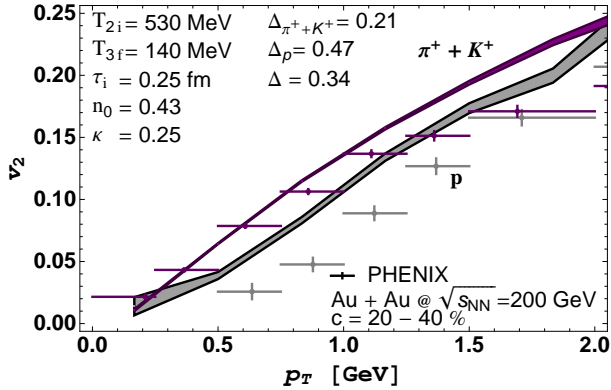}
\end{center}
\vspace{-3mm}
\caption{(Color online) The comparison of the model and experimental elliptic flow coefficient $v_2$ for pions+kaons and protons. The data are taken from Ref. \cite{Adler:2003kt}. The model parameters the same as in Fig. \ref{fig:ppt}. The bands indicate the error of the Monte-Carlo calculations.}
\label{fig:v2}
\end{figure}
\begin{figure}[t]
\begin{center}
\includegraphics[angle=0,width=0.45 \textwidth]{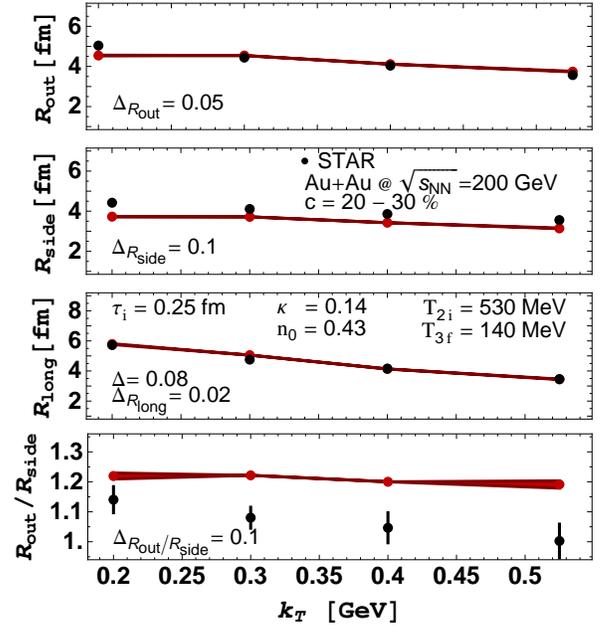}
\end{center}
\vspace{-3mm}
\caption{(Color online) The comparison of the model and experimental results for the pionic HBT radii. The data are taken from Ref. \cite{Adams:2004yc}. The model parameters the same as in Fig. \ref{fig:ppt}. }
\label{fig:hbt}
\end{figure}

\section{Comparison with data}
\label{sect:results}

The methods presented in the previous Sections delivered us the optimal parameters for fitting the RHIC Au+Au data for the centrality class 20-30\%. In this Section we present the direct comparison of our model results with the data. In Fig.~\ref{fig:ppt} we show the model and experimental transverse-momentum spectra of pions, kaons, and protons. The average agreement between the model and the PHENIX data ~\cite{Adler:2003cb} is quite satisfactory (\mbox{$\Delta$ = 13\%} for pions, 7\% for kaons, and 28\% for protons).

Figure \ref{fig:v2} shows our results for the elliptic flow compared with the PHENIX experimental results from Ref. \cite{Adler:2003kt}. We are able to reproduce very well the experimental pion+kaon elliptic flow (at the level of 12\%), however the model proton $v_2$ is too large by about 50\%. The large theoretical values of the proton elliptic flow are connected with the lack of the hadronic interactions in the final state or with the neglecting the viscous effects (as pointed out recently in Ref. \cite{Bozek:2009dw}). 

In Fig. \ref{fig:hbt} we show our results describing the HBT radii and compare them with the STAR results \cite{Adams:2004yc}. We observe a very good agreement between the model and the experimental data. The radii are reproduced with the accuracy better than 10\%.

We note that a similar analysis may be performed for the central collisions. With a slightly changed values of the input parameters we are able to reproduce the experimental data with a similar accuracy as in the case of the non-central collisions.

\section{Discussion and conclusions}
\label{sect:conclusions}

The main conclusion of our analysis is that it is possible to obtain a satisfactory description of the soft hadronic RHIC data in a hydrodynamic model that avoids difficulties connected with the concept of very fast 3D thermalization. In our model, the initial evolution of matter is described with the help of the transverse hydrodynamics that uses the assumption of the transverse thermalization only. The initial thermalization of the transverse degrees of freedom may be naturally explained in the string or color-flux-tube models. By combining the transverse hydrodynamics with the standard perfect-fluid hydrodynamics we include in the schematic way the effects of full 3D thermalization. 

We emphasize that our approach goes beyond the applications of viscous hydrodynamics, since the large initial anisotropy of pressure may be explained only by introducing a very large shear viscosity, see Eq. (\ref{emt_visc}),  which makes the very concept of the kinetic coefficients questionable. 

Our model reproduces well the HBT radii indicating that the two main problems of the soft physics at RHIC may be circumvented. The remaining issue of the large elliptic flow of protons may be very likely explained by the inclusion of the shear and bulk viscosity in the standard hydrodynamic stage or by the inclusion of the hadronic rescattering at freeze-out.  The results for other centralities will be published elsewhere \cite{Radek}.

In our earlier work \cite{Ryblewski:2009hm,Florkowski:2009wb} we considered a physical scenario where the transverse hydrodynamics was followed by a sudden isotropization and freeze-out. The main difference between the approach used in \cite{Ryblewski:2009hm,Florkowski:2009wb} and the present model is that we include now an extended phase governed by the equations of perfect-fluid hydrodynamics. The results of the two approaches are, however, quite similar (the approach with sudden isotropization and freeze-out gives slightly smaller HBT radii, on the other hand, their $k_T$ dependence is reproduced very well yielding a very good description of the ratio $R_{\rm out}/R_{\rm side}$). The similarities between our present results and the results of Refs. \cite{Ryblewski:2009hm,Florkowski:2009wb} indicate that the details of modeling the 2D $\to$ 3D transition are not very much important. The important point is that the matter should thermalize before freeze-out. Of course, the more realistic 2D $\to$ 3D switching may take place under different circumstances than those discussed in this paper or in Refs. \cite{Ryblewski:2009hm,Florkowski:2009wb}.

The present formalism may be generalized in many ways. In particular, the perfect-fluid stage may be replaced by the viscous hydrodynamics. The inclusion of the viscous effects is also possible in the transverse hydrodynamics. This opens new possibilities for the applications of the concept of transverse thermalization and for studies of the effects of anisotropic early dynamics.

\section{Acknowledgments} 

The authors thank Mikolaj Chojnacki and Adam Kisiel for computer assistance and discussions. This work was supported in part by the MNiSW grants No. N N202 288638 and N N202 263438.


\end{document}